# Angle resolved Photoemission from Cu single crystals; Known Facts and a few Surprises about the Photoemission Process

F. Roth[1], C. Lupulescu[2], E. Darlatt[3], A. Gottwald[3], W. Eberhardt[1,2]

[1] Center for Free Electron Laser Science (CFEL), DESY, Notkestr. 85, D-22607 Hamburg, Germany
[2] Inst. for Optics and Atomic Physics, TU Berlin, Strasse des 17. Juni 135, D-10623 Berlin, Germany
[3] Physikalisch Technische Bundesanstalt (PTB), Abbestr. 2-12, D-10587 Berlin, Germany

We present angle resolved photoemission spectra for Cu(100) and Cu(111) singly crystals in normal emission geometry, taken at tightly spaced intervals for photon energies between 8 eV and 150 eV. This systematic collection of spectra gives unprecedented insight into the influence of the final states to the photoemission process as well as the band structure and lifetimes of highly excited electrons in Cu.

**Introduction**

Angle resolved photoemission has been developed since 35 years ago as a method to investigate the band structure of the electronic states in solids and at surfaces (1-5). With the use of a continuously tunable excitation source, such as synchrotron radiation, it has become the experimental method of choice, when the momentum resolved energy band structure in solids, including their surfaces, are to be investigated in detail.

Initially all the spectrometers used in the experiments were designed by the individual groups interested in this experimental field, whereas by now there are several commercial instruments available (6, 7), which offer an over time constantly increased performance and resolution down to about 1 meV. Simultaneously to the instrumentation development, also the theoretical description of the process has undergone a continuous improvement. From the original three step model (8) to the understanding that the final state of the electron that is detected corresponds to the manifold of states that are observed in LEED (9-13). Taking the symmetry of the crystal and the detector position into account, powerful selection rules were derived which govern the photoemission process (14,15). Once band structure data were available, the theoretical descriptions were refined to overcome shortcomings or errors in the theoretical predictions. Initially this centered around the correct prediction of band gaps in semiconductors (16) and the prediction of metallic or nonmetallic behavior in the ground state. Even though these refinements are quite extensive and a lot of effort has been devoted to this, the correct treatment of electron correlation effects is still incomplete even if it is described by carefully chosen parameters (17). This leads to the well-established observation that the agreement between theory and experiments is much better for metals than for oxides.

Initially in experimental investigations most of the attention was devoted to gain information about the initial state of the photoemission process, in order to derive the occupied band structure of solids. Single crystalline Cu is and was the model system, where the technique of ARPES was tested and developed (19-26). The agreement with the calculated band structure was almost too perfect, even when the resolution of the experiment was improved (27-32). Applying the same methodology to Ni however, it immediately became apparent, that the situation is more complex (33). In the presence of strong electron correlation effects, the photoemission spectral function is observed, where the peak positions do not immediately reflect the ground state calculated bands anymore. This had been theoretically formulated by Hedin and Lundquist (34,35) and was quantitatively confirmed by studying the fairly simple band structure of an ordered CO adsorbate layer (36). This however means that now the relative peak intensities and line shapes have to be used in the interpretation, rather than the simple peak positions. Theory has advanced meanwhile to a state, where these spectral functions can be calculated under certain approximations and then compared with the actual photoemission data (17, 37-39). The agreement between theory and experiment for correlated magnetic solids however is far from perfect at present (40).

As a challenge to the state-of-the-art theory we here present photoemission data, taken in normal emission from Cu (111) and Cu(100) surfaces for photon energies varying between about 8 eV and 150 eV. The rationale behind that is that Cu should be much more simple to calculate than other transition metals. Cu presents (almost) no correlation effects and thus it should be more straightforward to compare the intensity variations of the spectral function as a function of the excitation energy with the theory. If theory fails to accurately reproduce these variations, then there is little hope that the much more complicated other transition metal systems can be represented accurately at the present state.

A second emphasis of this paper however is to get quantitative information on the electron self-energy or lifetime in the excited state. Concerning the electron self-energy and lifetime, theory is much more advanced than experiment (41-44). From early band structure measurements there are only a few data points on Cu available both from photoemission and inverse photoemission studies (24, 25, 45, 46). Beyond these studies, the real time resolved measurements using laser excited electrons and holes (47-49) only cover a very small energy range, due to the limitations in laser photon energies. Additionally surface state lifetimes have been measured by STS (Scanning Tunneling Spectroscopy) (50). Our goal here is to provide more systematic data by identifying direct transitions in the angle resolved (normal) emission from Cu (111) and Cu (100) oriented single crystals and to derive the lifetime width of the final state using the same formalism as previously established (24, 41, 43). Coupled with the group velocity of the electrons, this also gives an estimate of the effective escape depths.

**Experimental**

The photoelectron spectra were recorded with the SCIENTA R4000 hemispherical electron spectrometer in the iDEEAA end station (51) at Berlin's newest synchrotron radiation facility: Metrology Light Source (MLS), located in the Willy-Wien-Laboratorium of the Physikalisch-Technische Bundesanstalt (52). The high degree of p-polarization (99.5 %) was delivered by the IDB undulator beamline. Hereby the monochromator combines normal incidence (NI) with grazing incidence (GI) geometries and allows for monochromatic radiation from approximately 1.5 eV–10 eV in the NI mode and 10 eV–280 eV in the GI mode. The flux is about of $10^{12}$ photons/s at a 100 mA ring current and the resolving power is better than 900 (GI mode) and 2500 (NI mode). The spot size on the sample is about 1.7 mm horizontal and 0.1 mm vertical. The spectra were taken with the electron spectrometer at a pass energy of $E_{pass}$ = 20 eV, resulting in an electron energy resolution of 15 meV. The angular resolution was chosen to be ± 0.14 degrees by integrating the signal over the appropriate number of channels on the detector.

**Sample preparation and orientation**

The samples were prepared by several sputtering and annealing cycles and the cleanliness and order of the surface was verified by measuring the dispersion of the surface state (on Cu (111)). The Scienta electron spectrometer is mounted such that it takes a full angular spectrum of ± 7 degrees in the horizontal direction, whereas in the perpendicular direction we had to rely upon the mechanical mounting of the crystal and the manipulator (51). Measuring the surface state dispersion curves of the Cu (111) crystal surface we estimate the vertical angle to be aligned with an accuracy better than 0.5 degrees. Since identical mounting and sample holders were used for the Cu(100) crystal this estimate also holds for this surface. The angle of incidence of the light was 45° and the polarization was p-polarized in the plane of detection of the analyzer. This polarization enables us to observe all direct transitions possible in normal emission geometry as postulated by the selection rules (14, 15).

**Results and discussion I — band structure; initial and final states**

The normal emission spectra of Cu(100) and Cu(111) are shown in Fig. 1 for a series of photon energies between 8 eV and 150 eV. The intensity is presented as a false color plot. The color scale is linear and the gray horizontal lines indicate separate data sets. For each of the panels separated by the gray lines the spectra were taken consecutively during one injection. The spectra were normalized by integrating the emission current from the Au metallic coating of the last mirror in front of the sample to a preset value. This emission current was not scaled to reflect the incoming photon flux. This enables us to show these spectra using one color scale over the entire range of photon energies. At the boundaries between the data sets the integrated photoemission intensity was carefully matched.

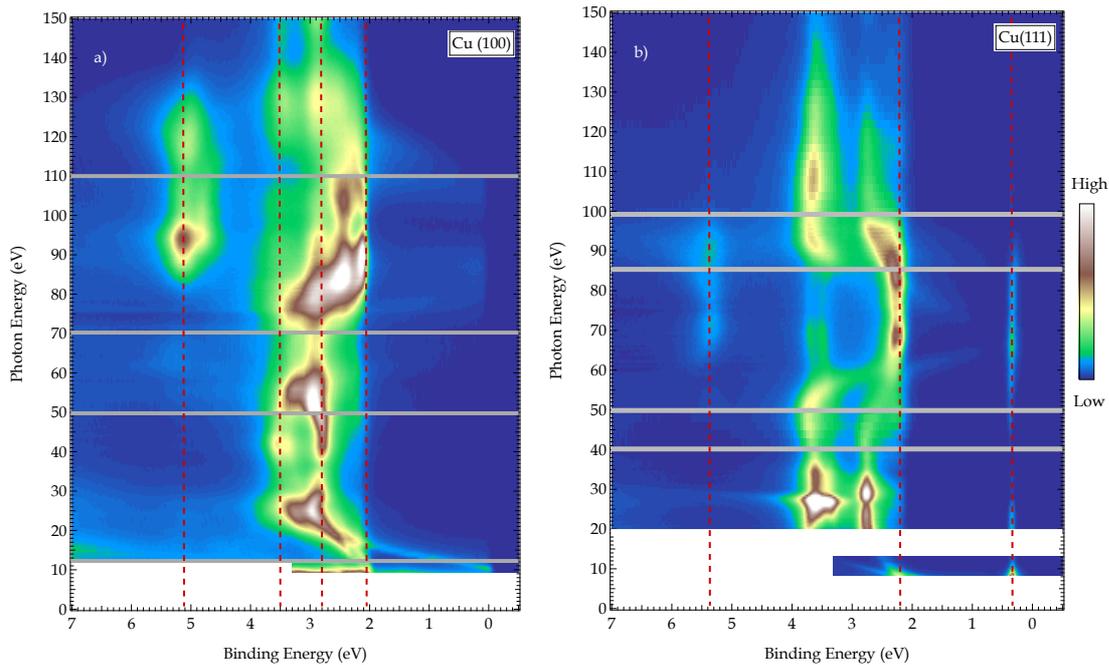

Fig. 1 (a) Left side: normal emission photoemission from the (100) surface, displaying the band dispersion along the $\Delta$-axis; (b) Right side: normal emission from the (111) surface displaying the band dispersion along the $\Lambda$-axis. The gray horizontal lines denote the regions of spectra measured during one injection. Moreover, the red dashed, vertical lines indicate the position of the cuts shown in the second part of the paper.

In order to facilitate the discussion, we show the bandstructure of Cu as originally calculated by G.A. Burdick (54) in Fig. 2. As the excitation energy is varied, the direct transitions shown in the spectra in Fig.1 occur at different points of the $\Delta$ ($\Lambda$) axis of the bulk Brillouin zone, as postulated by (parallel) momentum conservation.

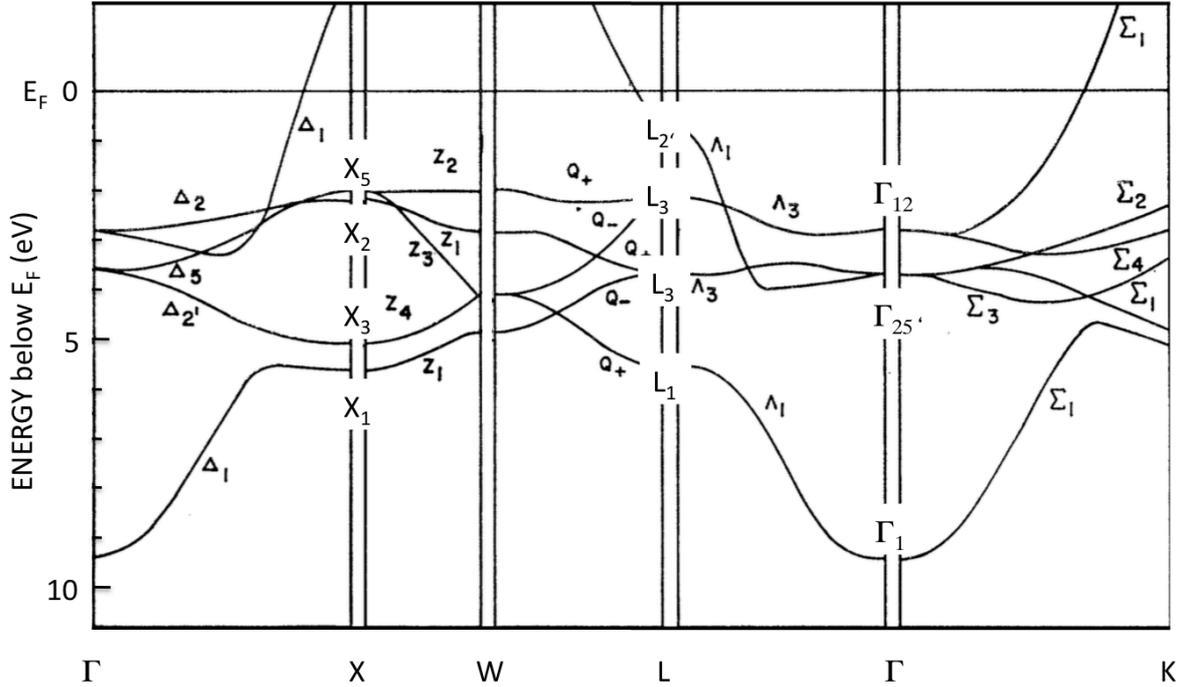

Fig. 2 Calculated bandstructure of Cu and band symmetries in a non-relativistic notation adapted from (54).

Based upon a simple interpolation and free electron like final state approximation with an inner potential of 11 eV (45), we expect to observe the emission from the Γ-point of the bulk band structure for photon energies around 41 eV in Fig. 1 a) and 30 eV and 135 eV in Fig. 1 b). The X-point emission is observed for photon energies of about 10 eV and 100 eV in Fig. 1 a), and correspondingly the L-point emission is observed around 74 eV photon energy in Fig. 1 b)

We immediately notice that this simple final state approximation does not describe the emission features seen in Fig. 1 completely. While the emission from the L-point is observed at photon energies around 70 eV in Fig. 1b, the reappearance of the same band features at photon energies around 85 eV are not described by this model. This refers to both the top of the d-band emission, at a binding energy around 2.20 eV, which corresponds to the $L_3$-point of the bulk band structure as well as the feature around a binding energy of 5.4 eV which corresponds to transitions involving the highest binding energy $\Lambda_1$ band. The exact location of this transition within the Brillouin zone is not quite as well determined as the top of the d-band, since this $\Lambda_1$ band has fairly flat regions throughout the Brillouin zone which correspondingly have a high density of states. The topmost point of the d-band (in this crystal orientation) however is only reached at or close to the L-point.

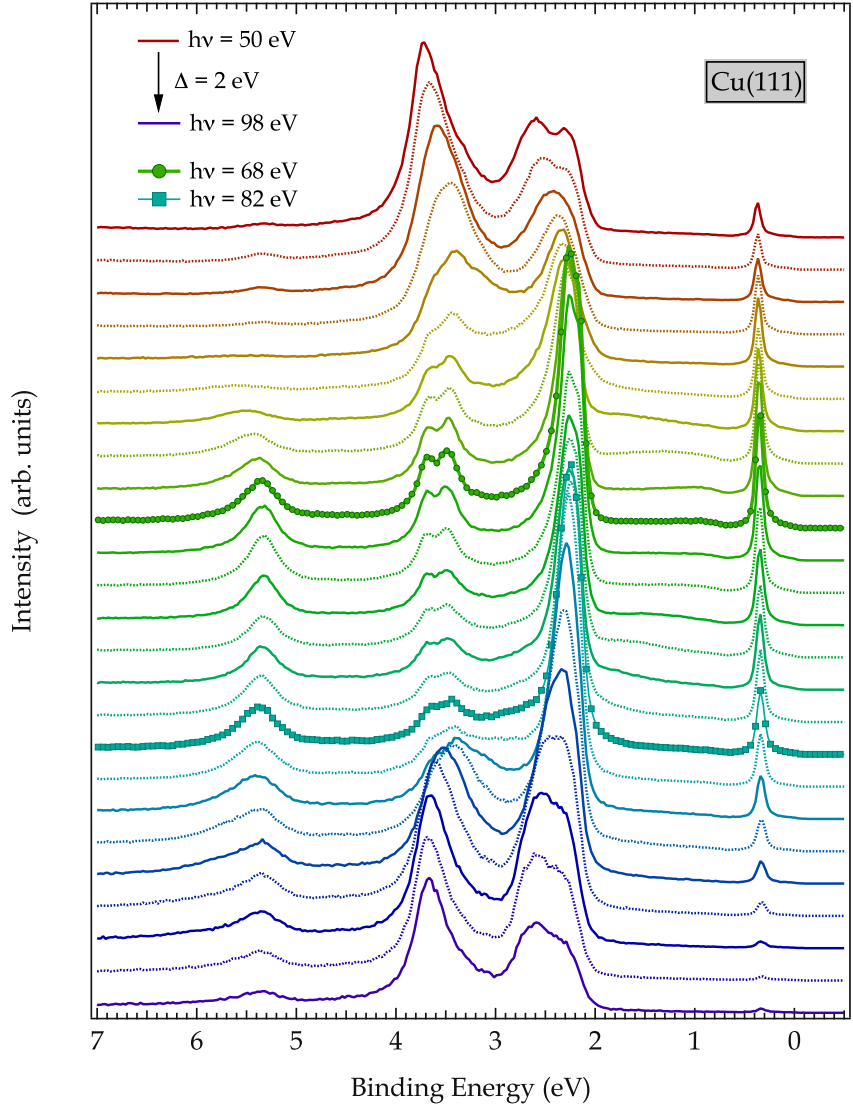

Fig. 3: Normal emission energy distributions obtained from Cu (111) at photon energies $50 \leq h\nu \leq 98$ eV using p-polarized light. The highlighted spectra at photon energies of 68 eV and 82 eV are discussed in the text.

In order to investigate this in more detail we present a few selected spectra showing transitions close to the critical points of the bulk band structure. Fig. 3 shows the emission for a long sequence of photon energies ranging from 50 eV to 98 eV. The analogous information is contained also in the false color plot in Fig. 1 b), but here the actual spectra are shown. The lowest binding energy peak, at $E_B = 0.45$ eV corresponds to the emission of the well known surface state in the center of the (111) surface Brillouin zone (SBZ). The next peak, which is composed of two emission features at 2.1. and 2.3 eV binding energy, originates from the uppermost $\Lambda_3$-band at the L-point, which is split by spin orbit interaction into two components. Analogously the two features at a binding energy of 3.6 eV and 3.8 eV are due to the emission of the lower $\Lambda_3$ band at the L-point. Finally the feature at a binding energy of 5.4 eV is due to the emission from the lowest $\Lambda_1$-band near the L-point.

The intensity maxima for the various emission features around or slightly below a photon energy of 70 eV are readily explained within the free electron final state approach. As a first approximation this works quite well. The additional maxima at lower energy or above 80 eV photon energy however do not fit into this picture. Before we discuss this any further however, we want to look at emission from the other critical points of the band structure, $\Gamma$ and X.

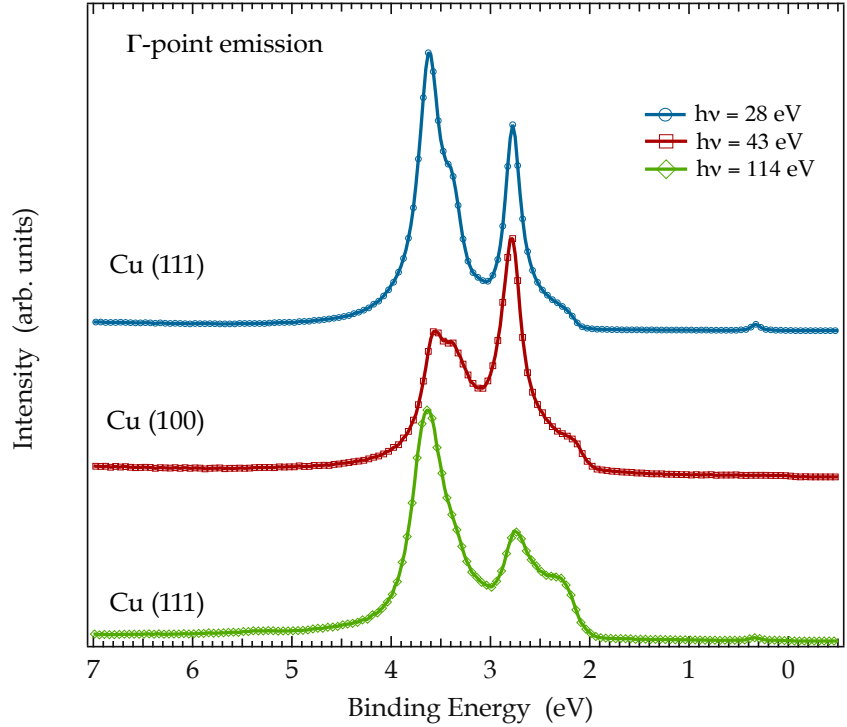

Fig. 4: Normal emission energy distributions obtained from Cu (111) at photon energies of 28 eV (top, blue circles) and 114 eV (bottom, green diamonds) and from Cu (100) at photon energy of 43 eV (middle, red squares) using p-polarized light.

The emission from the Γ-point is observed for both crystal orientations. Fig. 4 shows spectra taken at a photon energy of 28 eV and 114 eV from Cu(111) and a spectrum taken at 43 eV excitation energy from Cu(100). The two lower energy values fit quite well to a free electron final state, as discussed above. The curves were chosen by monitoring the dispersion of the bands rather than by using the free electron final state approximation. We note here that the high energy final state Γ-point is about 20 eV lower than the free electron energy estimate. In the (100) orientation the transitions are located at the Γ-point, when the $\Delta_5$ band reaches its maximum value of binding energy at about 3.7 eV. Analogously, in the (111) orientation the Γ-point is reached, when the upper $\Lambda_3$-band has a maximum in the binding energy at about 2.8 eV. The assignment of unique transitions in the Brillouin zone based upon observations for different crystal surface orientations and at well-defined emission angles has been used quite early on (23, 55, 56). This procedure was named triangulation method. In the spectra shown in Fig. 4 the top d-band emission around 2.8 eV binding energy is observed as a narrow single peak, whereas the second, higher energy d-band clearly exhibits two components at 3.4 and 3.7 eV binding energy due to spin orbit splitting. While the intensities of the direct transitions vary considerably for these spectra, the assignment is supported by the fact that the same peak positions are observed for all these spectra. The existence of the shoulder at binding energies around 2.3 eV points to the fact that other states are contributing to the photoemission process, which do not fit into the simple free electron final state approximation. This shoulder could be due to 'Umklapp' scattering or caused by coupling of high initial density of states regions to evanescent electron final states in the surface region of the crystal. It would be a challenge to theory to see, whether all these features, including the changes in the relative transition strengths for the different photon energies and sample orientations, can be correctly reproduced by present, state of the art calculations.

The emission from the X-point, on the other hand, is only observed in normal emission from Cu(100). In the energy range we are presenting here the X-point is reached twice within the free electron final state approximation. Once at or below 10 eV excitation energy and the other one taken at a photon energy of about 110 eV. The latter assignment is corroborated by the series of spectra presented in Fig. 5, where it can be seen that only in a narrow region around 110 eV excitation energy the dominant features observed are the topmost d-band emission between 2.0 and 2.4 eV binding energy in conjunction with the s-p band emission at 5.2 eV binding energy. Interestingly at only slightly lower photon energies the spin-orbit splitting of the topmost d-band into three components is clearly recognizable in the spectra, while the $\Delta_1$ band that crosses the Fermi level near the X-point is only faintly recognizable. This band is better visible in the false color representation in Fig. 1 a). Additionally there is always a weak emission seen at binding energies between 3.5 and 4 eV. This cannot originate from transitions near the X-point, but, as the false color representation in Fig. 1 shows, emission at this binding energy is always observed almost independent of the excitation energy. The higher binding energy part of the d-band has a fairly flat dispersion over large ranges of the Brillouin zone and accordingly a large density of states is found at binding energies between 3.5 and 4 eV. The emission from these states has to originate from non free electron like final states contributions to the photoemission process.

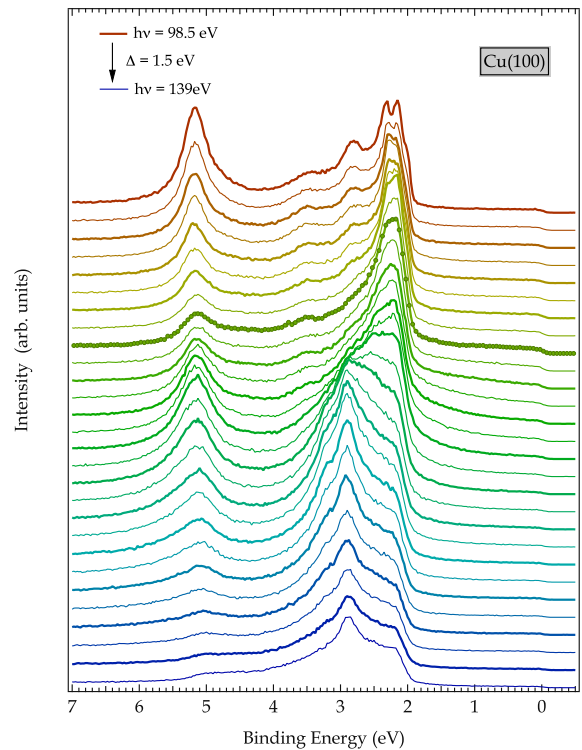

Fig. 5 Series of normal emission spectra from Cu(100) taken for photon energies between 98.5 eV and 139 eV. The highlighted spectrum at excitation energy of 110 eV is discussed in the text.

Moreover and very much to our surprise, the spectra shown in Fig. 5 also display pronounced variations for relatively small changes in excitation energy. At an excitation energy of about 110 eV (kinetic energy of 100 eV), a change of the excitation energy by 10 eV corresponds to a change in normal momentum of 0.26 Å$^{-1}$ which is only about 15% of the reduced momentum along the $\Delta$-axis ($\Gamma$X=1.74 Å$^{-1}$). The EDC's displayed in Fig. 5 are taken for increments of 1.5 eV in photon energy. Accordingly a 10.5 eV change occurs over 7 curves. Or in another reference scale the all spectra shown in Fig. 5 cover the normal momentum range from X – 18%$\Gamma$X to X + 40%$\Gamma$X. In the extended zone scheme we are probing a region around a normal momentum of 5.2 Å$^{-1}$, which corresponds to the second X-point in the direction of the surface normal. Inspecting the calculated band structure (see Fig. 2), very little dispersion should be observed in the d-bands for an approximate 30% change in reduced momentum along the Δ-axis around the X-point. Especially we do not expect any emission from a state at a binding energy around 2.8 eV, which is indicative for transitions more than 45% ΓX-momentum distance away from the X-point in this sample orientation. However within a 10 eV change in excitation energy around the X-point, a well distinguished feature at a binding energy around 2.8 eV is observed. The finite angular acceptance and possible misorientation by 0.5 degrees (see above) amount to much smaller uncertainties in the momentum determination of less than ± 0.1 Å$^{-1}$. Final state symmetry changes around the gap at a critical point in the bandstructure could in principle account for drastic changes in the observed transition intensities. However, as was pointed out quite early (14,15), in normal emission geometry the wavefunction of the outgoing electron in vacuum has to be totally symmetric under all symmetry operations. Depending on the surface orientation this corresponds to either $\Delta_1$ or $\Lambda_1$ symmetry. This puts quite severe constraints on the wavefunction matching in the surface region. Moreover, the feature at a binding energy of 2.8 eV is observed with comparable intensity upon approaching the X-point from either direction.

On the positive side this clearly indicates that the final state structure has a pronounced effect on the measurements even at kinetic energies more than 100 eV above $E_F$. The spectra unambiguously show transitions which cannot be assigned by assuming a free electron final state. The structure of the electronic states inside the crystal or in the surface region introduces pronounced contributions to the photoemission process even at these fairly high energies. This has to be taken into account, whenever spectral intensities are calculated. The detailed coupling of the initial state energy bands inside the crystal to all existing inverse LEED states of the outgoing electron that is actually detected in vacuum is absolutely crucial for a proper description of the photoemission features. This not only concerns the spectral intensity of the major features, but more distinctly the appearance of peaks and shoulders, which are associated with emission from points in the Brillouin zone, which cannot be assigned within a simple free electron like final state approximation. We want to point out that this observation is not restricted to the (100) surface orientation of the sample. The spectra in the direction of the (111) surface normal show quite similar effects and this information is contained in the false color plots presented above in Fig. 1a and b.

From the early days angle resolved photoemission from Cu has been the 'gold standard' to develop the method and to actually experimentally determine the dispersion relation of the electronic bands in a solid and on its surface (19, 20, 23, 24, 25). Soon it was recognized that this experimental method works exceptionally well for Cu, while for other transition metals, for example Ni, the agreement between the measured electronic states in photoemission and the calculated bandstructure was far from perfect (33). This was attributed to electron correlation effects. As soon as the outgoing electron interacts too strongly with the other electrons in the crystal, these interactions give rise to additional structures in the spectra and energy shifts of the observed features. The photoemission measurements actually reflect the spectral function of the emitted electron, which reflects not necessarily the ground state of the remaining system, but specifically also excited states to the degree they are populated in the photoemission process. For very simple systems, where only two different final states are involved, this can be corrected for already in the analysis of the experimental data (36), but for more complex system with multiple excited states the help of theory is required for a proper analysis and interpretation of the spectra.

Resonant photoemission, which can also lead to distort the spectral intensities and cause additional satellite structures is relatively weak in Cu. Furthermore the corresponding satellite structures are located at energies between 10 eV and 15 eV below $E_F$, well outside our range of observations (57). The core level excitation thresholds of the Cu 3p and 3s core electrons are located at 75.0 eV, 77.7 eV and 122.4 eV.

By now theory has been developed to take electron correlation effects into account and theoreticians engaged in calculating photoemission spectra rather than band structures. This has been tested in some detail for the first row transition metals Fe, Co, and Ni (40), but the agreement between the measured and calculated spectra was less than satisfactory. We hope the data presented here will be taken as a challenge to improve upon these theoretical calculations, in a case, where electron correlation effects are much less severe than for the magnetic transition metals.

**Results and discussion II — final state self energy and electron lifetime**

As it was demonstrated before (24, 45), measuring the intensity of a well known interband transition at a defined point of the band structure as a function of excitation energy allows to determine both the initial state hole and final state electron lifetime or self energy. These self-energies have been calculated for energies up to more than 100 eV, however experimental data is extremely scarce and limited to only low energies (42). For very low energies these lifetimes have been measured by time resolved two photon photoemission, but in these experiments the excitation energy is limited by the wavelength of the (optical) pump laser pulse. Also at the surface, image state lifetimes have been determined using this technique (47-50).

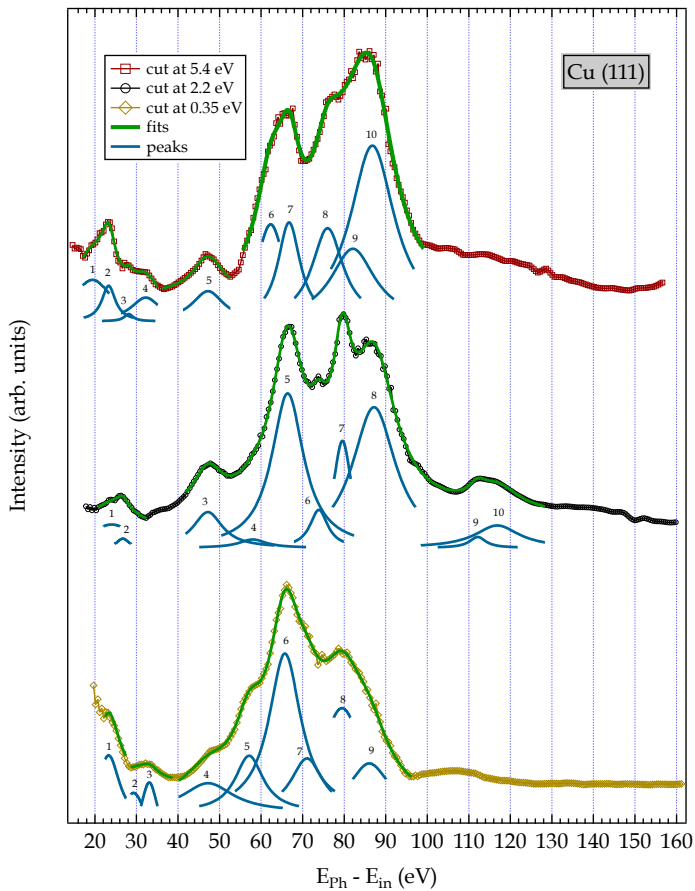

Fig. 6 Intensity of selected features in the EDC maps (Fig.1b) plotted as a function of the final state energy. These curves are obtained by cuts along vertical lines in Fig. 1b, i.e. at a constant initial state energy. Additionally, in order to account for the differences in initial state energies, the curves are plotted on the final energy scale ($E_{fin} = E_{ph} - E_{in}$). The bottom (light green) curve shows the emission profile of the well known surface state, whereas the middle (green) curve shows the emission from the $L_3$ point and the top (red) curve shows the intensity profile of transitions located at the $L_1$ point. The very complex shape of these curves is at certain energies locally approximated by fitting Lorentzians to small regions as indicated by the blue curves. The fit parameters (energy position, FWHM) are listed in Table 1.

The data shown in Fig. 1 additionally enable access to the determination of the electron and hole lifetimes. Using vertical cuts through these data, the intensity of certain transitions as a function of the photon energy can be mapped. The crucial point in the analysis is to be able to have a well-defined assignment for the transitions that are observed. Therefore lets start by evaluation the intensity of the surface state emission from the center of the Brillouin zone of the (111) surface. This has been published early on (25) but was not evaluated in detail. The green (bottom) curve in Fig. 6 shows this intensity variation for photon energies between 20 and 150 eV.

Fig. 6 also shows the intensity variations of other band features associated with the emission near the L-point of the bulk band structure for photon energies between 20 eV and 150 eV photon energy. The surface state emission intensity clearly shows a different profile than the other emission features. The surface state emission displays a clear maximum for excitation at a photon energy of 66 eV with two distinct shoulders at 58 eV and 80 eV photon energy. The maximum of this emission has been reported and discussed before (25). Interestingly both of these shoulders are also present in the earlier data, but were not discussed, probably because of the statistical uncertainties. The peak in the surface state emission at 66 eV photon energy is attributed to the fact, that the outgoing final state wave function has a wavelength, which corresponds to the lattice plane distance along the (111) crystal axis. The initial surface state wave function is distributed over about three to four lattice planes and the matching of the contributions of the emission from all these planes is optimal, when the final state has the appropriate wavelength (25).

While for the surface state the coupling is strongest for a photon energy of about 66 eV, the emission associated with the transitions from the upper and lower d-band at the L-point ($L_3$ and $L_1$ in non-relativistic denomination) displays a different intensity profile. Remarkable however is that the observed structures align fairly well according to the final state energy. This clearly indicates that the final states for these transitions are largely the same, whereas the relative coupling, as described by the transition matrix elements, changes. Moreover we have to keep in mind, that the surface state and the bulk features may couple differently to the available final states. For bulk states, the transitions predominantly are into bulk final states, which exist in the bulk solid and couple to inverse LEED states near the surface of the crystal. For the surface state there is an enhanced probability to couple directly to the inverse LEED states, even when these have highly damped evanescent amplitude in the direction perpendicular to the surface. In the energy range between 66 and 86 eV above $E_F$ at least three different final state bands exist at or near the L-point, which have a major (111) momentum component, i. e., they lead to photoemission along the surface normal of the (111) oriented crystal.

We have carried out a components analysis of the complex curve showing the intensity of certain initial state features in the photoemission process. Local maxima of these curves have been fitted by Lorentzians as indicated in the Fig. 6 by the numbered peaks. The rationale for this is that both the initial state as well as the final state of the transition are characterized by a lifetime or self-energy and the line shapes are Lorentzians, excluding other (i.e. phonon) broadening contributions (41). Starting with the surface state, the initial state lifetime broadening is on the order of about 24 meV as determined by scanning tunneling spectroscopy (STS) (50). Accordingly the peak width in the fits in Fig. 6 is essentially due to the final state liefetime or self-energy. This width increases from about 1 eV at lower energies (10 to 20 eV) to almost 10 eV around 70 eV and even more for higher energies. Table 1 lists the essential fit parameters such as peak position and width for the components used to model the complex function of intensity vs. excitation energy for all transitions observed for the (111) oriented crystal.

Obviously the angular (momentum) resolution and the slope (group velocity $\delta E/\delta k$) of the bands have to be accounted for in extracting the lifetimes, since they all contribute to the peak width observed for the transition, however the major factor responsible for the large increase in width with photon energy is by far the change in lifetime of the final state. The broadening due to the group velocity and finite angular resolution can be estimated by assuming a free electron like final state band. This actually results in an upper bound for this contribution to the linewidth, since the actual slope of the bands will generally be smaller than the free electron bands. This especially holds for transitions near critical points where the slope approaches zero. The slope of a free electron like final band is given by

$$\delta E/\delta k = 2E/k$$

Accordingly the broadening due to the finite angular resolution can be expressed as

$$\Delta E = 2E\, \Delta k\, /k = 2E \sin(\delta\Theta)$$

For an angular resolution of $\delta\Theta = \pm 0.14°$ this relates to $\Delta E = 0.01E$.

This means that the maximum linewidth contribution due to the finite angular resolution and group velocity of the final state bands is about 1% of the final state total energy. This is much smaller than other systematic error contributions resulting from the crude approximation of modeling the curves locally by a series of Lorentzians.

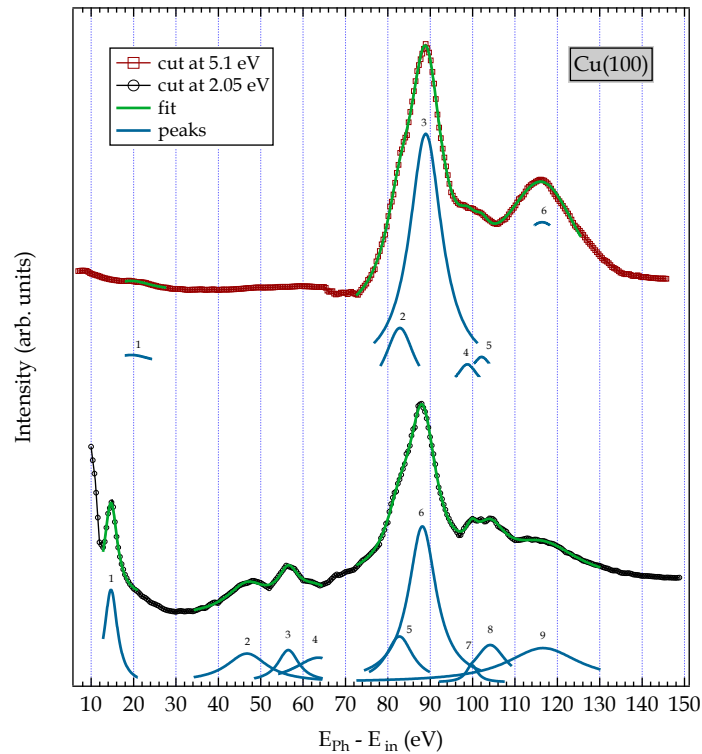

Fig 7. Emission intensity profiles for Cu (100) corresponding to emission from the top of the d-band ($X_5$, $E_{initial}$= 2.05 eV) and to emission from the highest binding energy band ($X_1$, $E_{initial}$= 5.1 eV) at the X-point.

Fig. 7 shows the corresponding emission intensity profiles for the (100) oriented crystal surface taken at an initial state energy of 2.05 eV, which corresponds to emission from the top of the d-band ($X_5$) at the X-point and at an initial state energy of 5.1 eV, which corresponds to the highest binding energy d-band emission at the X-point ($X_1$). The most prominent feature is observed at a final state energy of about 88 eV for both initial states. This is about 12 eV below the value determined by the admittedly crude approximation of free electron like final states.

Figs. 8 and 9 show emission profiles corresponding to initial state energies of 2.8 eV (Fig.8) and 3.5 eV (Fig. 9). Nominally these initial state energies correspond to the emission from the Γ-point of the band structure. However the d-band dispersion is quite flat, such that at these initial state energies the assignment is not uniquely derived. Moreover, in the free electron final state approximation, which works quite well, as discussed above, for the L-point and the X-point, the Γ-point is reached at a final state energy around 41 eV for a (100) oriented crystal. The strongest features observed in Figs. 8 and 9 are at quite different final state energies. Remarkeably, there are quite strong and prominent transitions observed at several final state energies in these profiles. Here we are also interested in getting lifetime or self-energy information for as many final states as possible and accordingly we include these transitions in the values listed in table 2. As the initial band dispersion in the d-bands is quite flat, the observed width of these structures is predominantly due to the self-energy/lifetime in the final state of the transition. Additionally the hole lifetime also contributes to these observed widths. However the hole lifetime is small, compared to the measured widths in the intensity profiles. According to published experimentally determined (49, 58) as well as theoretically calculated values (59,43) the hole lifetime broadening is about 70 meV at the top of the d-band and increases to 330 meV at the bottom of the d-band ($X_1$, $L_1$) due to an increasing d-d Auger decay rate for holes in the d-band with increasing binding energy. For a simple Fermi liquid this hole lifetime broadening would increase proportionally to $(E_h-E_F)^2$. Goldmann et al. (58) actually give an experimentally determined approximation for d-band holes in Cu as

$$\Gamma_h(\text{eV}) = 0.012\text{eV}^{-1}(\text{E}_h - \text{E}_F)^2 + 0.02\text{eV}$$

However we have to note here that the lifetimes determined by decoherence effects in fs laser photoemission (49) for the top of the d-band ($X_5$) are about a factor of 3-4 longer than the ones determined by photoemission (54).

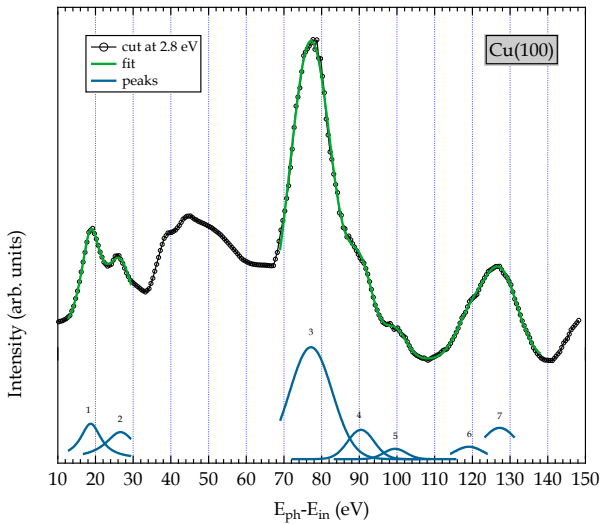

Fig. 8 Intensity profile of the direct interband transitions observed for an initial state energy of 2.8 eV for the Cu (100) oriented crystal

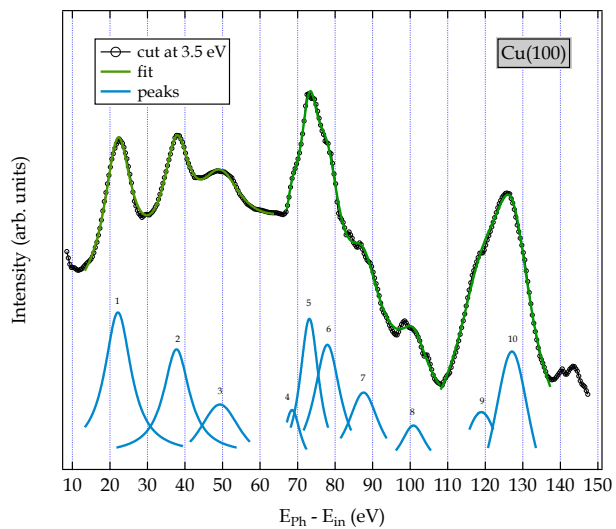

Fig. 9 Intensity profile of the direct interband transitions observed for an initial state energy of 3.5 eV for the Cu(100) oriented crystal

Nevertheless in comparison with the widths of the transitions observed in Figs. 6-9 the hole state width is negligible. The width of these transitions is largely determined by the final state lifetime width.

The overall magnitude of the intensity variations observed in Figs. 6-9 is also worth a discussion. While the concept of the inelastic mean free path for electrons has been generally accepted, the data presented her clearly show that direct transitions dominate the picture up to kinetic energies of at least 100 eV. This can be extracted from the intensity variation observed when a direct transition is encountered. As documented in Figs. 6-9 the photoemission intensity changes by up to an order of magnitude when a direct transition is encountered. These electrons assigned to an interband transition have obviously not undergone inelastic scattering and their contribution dominates the spectra, even at kinetic energies around 50 to 70 eV, where the inelastic scattering is supposed to be strongest and the inelastic path shortest.

To summarize these observations we have plotted the width of the observed transitions as a function of final state energy in Fig. 10. In this figure we have omitted a few of the peaks listed in table 1 and 2, where the width is extremely large. In some instances there are indications that several transitions are not resolved. Even though the values plotted in Fig. 10 display a lot of scatter, the overall trend indicates a monotonically increasing width of the final state lifetime broadening, but not necessarily a quadratic dependency as formulated earlier. There is no obvious dependence on crystal orientation, which is not unexpected.

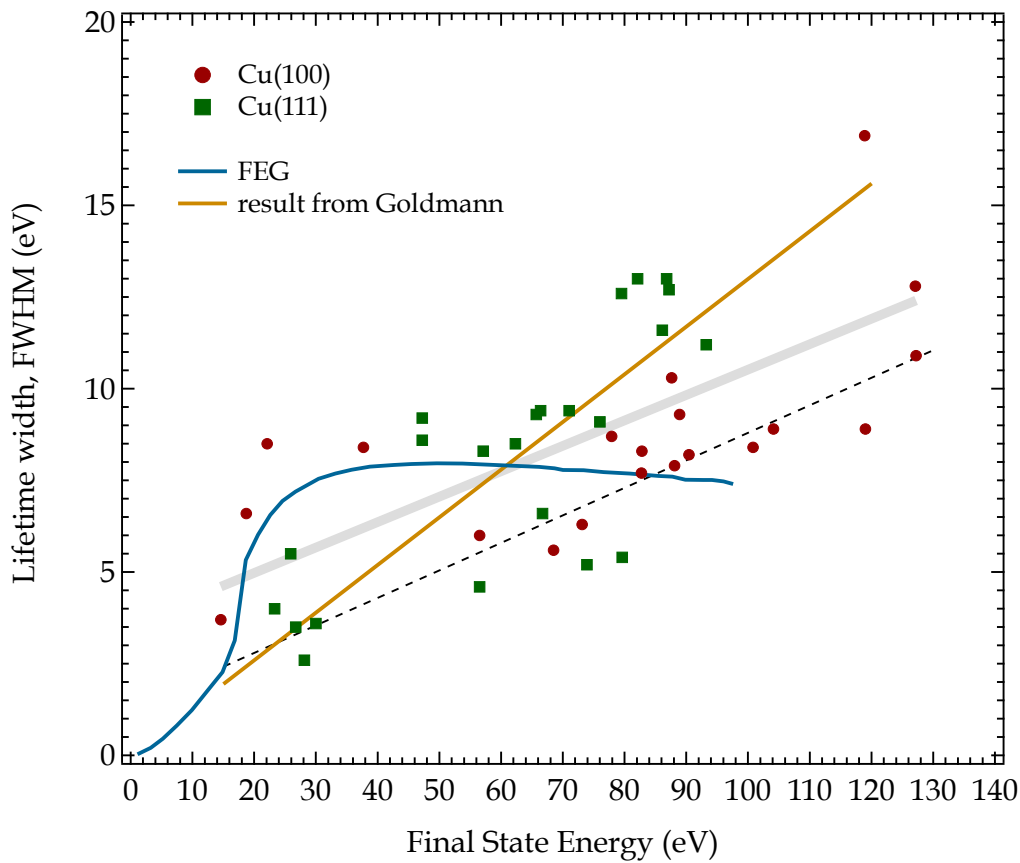

Fig.10 Lifetime width of the observed transitions as a function of the final state energy above $E_F$ for Cu(100) (red circles) and Cu(111) (green squares). The free-electron-gas model (blue line) was taken from Ref. (42) and the (grey) linear function is a guide to the eye. The yellow line shows the empirical estimate from Goldmann et al (46) and the dashed line is explained in the text.

Very low energy excited electrons within a few eV above $E_F$ display a very sharp drop in lifetime with increasing excitation energy and correspondingly a lifetime width which increases much stronger than linear (47-49). On the other hand, for the energy range from 20 eV to above 100 eV the lifetime $\tau$ increases much less rapidly. Goldmann et al. (46) empirically determined a linear rise approximated as $\tau^{-1} = 0.13 (E - E_F)$. This empirical relation is shown as a yellow solid line in Fig. 10. Also included in Fig. 10 is the lieftime as predicted by a free electron gas model (blue line) using the electron density of Cu as calculated by Echenique et al. (42).

We consider it appropriate to view the lifetime width data in Fig. 10 as an upper bound of the actual lifetime or self energy in the final state of the photoemission process. Other factors such as phonon scattering or unresolved multiple transitions might lead to a measured width which is larger than the lifetime width. Therefore it might be appropriate to concentrate on the sharpest transitions observed, rather than all transitions. This is suggested by the dashed line in Fig. 10, which follows the relation $\tau^{-1} = 1.3$ eV $+ 0.075 (E - E_F)$. There is no thorough mathematical or physical derivation for this — just an empirical estimate following the collection of data points near the lowest values. On the other hand, the lifetime as calculated for the free electron gas really does not reflect the behavior observed in our data.

We have to add here that a lifetime which decreases linear with energy from 20 eV to above 100 eV is not consistent with a minimum in the electron escape depth around 50 eV kinetic energy, since the approximate group velocity of the electrons increases proportional to $E^{1/2}$. This is another manifestation that the transitions which are observed in the unscattered electron signal indeed do reflect bulk interband transitions and are not limited or distorted by the electron escape depth.

**Conclusions**

We have presented angle resolved photoemission intensity maps taken in the direction of the surface normal of a Cu(111) and a Cu(100) oriented crystal. These are intended to serve as a well-defined test case for full photoemission calculations as they have become possible recently. The Cu initial state electronic structure has been extremely well characterized in the past, but the data presented here aim at defining the final states participating in the photoemission process.

The first and most important observation here is that up to at least 100 eV final state energy a multitude of clear and well-defined final states are observed in these photoemission spectra and even beyond these energies direct interband transitions are still detectable. The free electron like final state approximation works reasonably well as a first approximation, but on the other hand there are also other transitions, possibly due to Umklapp scattering, contributing to the spectra. Accordingly, when calculating photoemission spectra the full band structure has to be taken into account up to at least 100 eV photon energy.

On the other hand, this is also the manifestation that band structure mapping by angle resolved photoemission indeed works as pronounced by the protagonists of this technique. Inelastic electron scattering, the so-called inelastic mean free path, does not overpower these observations. Inelastic scattering processes lead to a general increase in the (unassignable) background of the spectra. As the intensity variations demonstrate, even at kinetic energies around 50 to 70 eV, the direct transition related signal is a factor of 5-10 larger than the background. As long as these well-defined transitions are observed, these occur between initial states and final states inside the crystal and at its surface. Inside the crystal these are the final state bands of the band structure and in the surface region there are additional surface states and resonances as well as evanescent free electron states from the vacuum.

## Acknowledgment

We thank the staff of the MLS, especially to H. Kaser, for the experimental and technical support. This research is funded by the Helmholtz Association of Research Centers in Germany within the program 'Structure of Matter'.

**Cu (111)**

| | Peak | Location (eV) | FWHM (eV) |
|---|---|---|---|
| **cut at 5.4 eV** | 1 | 19,415 | 12,602 |
| | 2 | 23,282 | 3,965 |
| | 3 | 28,122 | 2,588 |
| | 4 | 32,225 | 8,502 |
| | 5 | 47,194 | 9,189 |
| | 6 | 62,287 | 8,503 |
| | 7 | 66,714 | 6,563 |
| | 8 | 75,992 | 9,135 |
| | 9 | 82,079 | 12,993 |
| | 10 | 86,774 | 12,998 |
| **cut at 2.2 eV** | 1 | 24,033 | 13,659 |
| | 2 | 26,701 | 3,486 |
| | 3 | 47,171 | 8,581 |
| | 4 | 57,969 | 8,012 |
| | 5 | 66,387 | 9,393 |
| | 6 | 73,891 | 5,219 |
| | 7 | 79,562 | 5,375 |
| | 8 | 87,168 | 12,727 |
| | 9 | 112,120 | 4,738 |
| | 10 | 116,820 | 12,994 |
| **cut at 2.8 eV** | 1 | 25,888 | 5,486 |
| | 2 | 30,048 | 3,611 |
| | 3 | 36,894 | 7,932 |
| | 4 | 42,881 | 9,000 |
| | 5 | 56,454 | 4,615 |
| | 6 | 93,152 | 11,175 |
| | 7 | 116,15 | 17,691 |
| | 8 | 131,02 | 6,474 |
| **cut at 0.35 eV** | 1 | 22,977 | 6,499 |
| | 2 | 29,278 | 8,844 |
| | 3 | 33,062 | 5,437 |
| | 4 | 47,218 | 14,725 |
| | 5 | 57,107 | 8,269 |
| | 6 | 65,720 | 9,350 |
| | 7 | 71,031 | 9,389 |
| | 8 | 79,474 | 12,624 |
| | 9 | 86,054 | 11,577 |

Table 1: Fit parameters for all transitions observed for Cu (111) oriented crystal (see Fig. 6). These Lorentzians serve as local approximations to the observed intensity profiles and are included as blue curves in Fig. 6

## Cu (100)

| | Peak | Location (eV) | FWHM (eV) |
|---|---|---|---|
| **cut at 5.1 eV** | 1 | 19,388 | 11,381 |
| | 2 | 82,815 | 8,366 |
| | 3 | 88,919 | 9,276 |
| | 4 | 98,750 | 5,901 |
| | 5 | 102,120 | 4,935 |
| | 6 | 116,390 | 24,461 |
| **cut at 2.05 eV** | 1 | 14,646 | 3,663 |
| | 2 | 46,720 | 11,719 |
| | 3 | 56,489 | 6,059 |
| | 4 | 63,594 | 13,692 |
| | 5 | 82,716 | 7,691 |
| | 6 | 88,130 | 7,924 |
| | 7 | 99,780 | 3,302 |
| | 8 | 104,140 | 8,943 |
| | 9 | 116,640 | 20,928 |
| **cut at 2.8 eV** | 1 | 18,697 | 6,646 |
| | 2 | 26,600 | 9,617 |
| | 3 | 77,162 | 13,216 |
| | 4 | 90,402 | 8,195 |
| | 5 | 99,470 | 7,000 |
| | 6 | 119,050 | 8,913 |
| | 7 | 127,200 | 10,898 |
| **cut at 3.5 eV** | 1 | 22,111 | 8,553 |
| | 2 | 37,722 | 8,391 |
| | 3 | 49,290 | 13,979 |
| | 4 | 68,525 | 5,605 |
| | 5 | 73,116 | 6,276 |
| | 6 | 77,909 | 8,699 |
| | 7 | 87,562 | 10,329 |
| | 8 | 100,830 | 8,402 |
| | 9 | 118,930 | 16,911 |
| | 10 | 127,090 | 12,800 |

Table 2: Fit parameters for all transitions observed for Cu (100) oriented crystal (see Fig. 7 – Fig. 9). These Lorentzians serve as local approximations to the observed intensity profiles and are included as blue curves in Figs. 7-9